\documentstyle[prl,aps,epsbox]{revtex}
\newtheorem{theorem}{Theorem}

\newtheorem{lemma}[theorem]{Lemma}
\newtheorem{corollary}[theorem]{Corollary}

\begin{document}
\draft


\title{
Strength of interaction for information distribution
}

\author{Takayuki Miyadera$\ ^*$
and Hideki Imai$\ ^{*,\dagger}$
}

\address{$\ ^*$
Research Center for Information Security (RCIS),\\
National Institute of Advanced Industrial 
Science and Technology (AIST).\\
Daibiru building 1102,
Sotokanda, Chiyoda-ku, Tokyo, 101-0021, Japan.
\\
(e-mail: miyadera-takayuki@aist.go.jp)
\\
$\ ^{\dagger}$
Graduate School of Science and Engineering,
\\
Chuo University. \\
1-13-27 Kasuga, Bunkyo-ku, Tokyo 112-8551, Japan .
}
%

\maketitle
\begin{abstract}
Let us consider two quantum systems: system $A$ and system $B$. 
Suppose that a classical information is encoded to quantum states of the 
system $A$ and we distribute this information to both systems by making 
them interact with each other. 
We show that it is impossible 
to achieve this goal perfectly if the strength of 
interaction between the quantum systems is smaller than a 
quantity that is determined by noncommutativity between 
a Hamiltonian of the system $A$ and the states (density operators) 
used for the information 
encoding. 
It is a consequence of a generalized Winger-Araki-Yanase theorem which 
enables us to treat conserved quantities 
other than additive ones. 
\end{abstract}
Let us consider two quantum systems, system $A$ and system $B$. 
Suppose that a (classical) bit is encoded to a pair of quantum states of 
the system $A$. To distribute (or broadcast) the information to 
both systems, one needs interaction between them. How strong the 
interaction should be? 
\par
Let us begin with a detailed explanation of the problem.
The system $A$ (resp. $B$) 
is described by a Hilbert space, ${\cal H}_A$ (resp. ${\cal H}_B$).
A classical information, $0$ or $1$, is encoded to 
a pair of pure distinguishable quantum states 
of the system $A$. $0$ is encoded to 
$|\psi_0\rangle \langle \psi_0|$, and $1$ is encoded to 
$|\psi_1\rangle\langle \psi_1|$, 
where $|\psi_0\rangle$ and $|\psi_1\rangle$ are the normalized vectors in 
${\cal H}_A$. We assume that the vectors $|\psi_0\rangle$ and 
$|\psi_1\rangle$ are orthogonal with each other.
The composite system, ${\cal H}_A\otimes {\cal H}_B$, is 
assumed to be a closed system. The time evolution of the closed system is 
determined by a Hamiltonian, $H:=
H_A+H_B+H_{int}$, where $H_A$ (resp. $H_B$)
is an operator acting only on ${\cal H}_A$ (resp. ${\cal H}_B$), and 
$H_{int}$ is an interaction term. Before the interaction, 
the system $B$ is assumed to be in a state $\sigma$ 
independent of the state of the system $A$. 
Our purpose is broadcasting the classical information to 
both systems. That is, states after the interaction should be 
perfectly distinguishable on both systems. Let us write 
$\rho_0$ and $\rho_1$ the states of the composite 
system after the interaction. 
If we put $T$ a time interval for the interaction, 
they can be written as 
$\rho_j:=U(|\psi_j\rangle \langle \psi_j|
 \otimes \sigma)U^*$ for $j=0,1$ with
$U:=e^{-iHT}$. 
Their restriction to the system $A$ (resp. $B$) defines
$\rho_j^A$ (resp. $\rho^B_j$) for $j=0,1$. 
To discuss distinguishability, we make use of 
a measure called fidelity\cite{Uhlmann,Jozsa}. The fidelity between 
two states $\sigma_0$ and $\sigma_1$ is defined 
as $F(\sigma_0,\sigma_1):=\mbox{tr}(\sqrt{\sigma_0^{1/2}\sigma_1
\sigma_0^{1/2}})$ which takes $1$ iff the states coincide with 
each other and 
takes a smaller nonnegative value 
as they are more distinguishable. 
The following lemma proved by \cite{FuchsCaves,BCFJS} is not only 
useful but also justifies that the fidelity indeed represents 
distinguishability of states.
\begin{lemma}\label{lemma1}
The fidelity equals the
minimum overlap of the 
square root of coefficient between 
two probability distributions $p_0$ and $p_1$:
\begin{eqnarray*}
F(\rho_0,\rho_1)=\min_{\{E_{\alpha}\}:POVM}\sum_{\alpha}
\sqrt{p_0(\alpha)p_1(\alpha)},
\end{eqnarray*} 
where $p_0$ and $p_1$ are defined by 
$p_0(\alpha)=\mbox{tr}(\rho_0 E_{\alpha})$ 
and $p_1(\alpha)=\mbox{tr}(\rho_1 E_{\alpha})$. 
The minimum is taken over all the possible positive 
operator valued measures (POVMs), where a POVM $\{E_{\alpha}\}$ is 
a family of the positive operators satisfying 
$\sum_{\alpha}E_{\alpha}={\bf 1}$. 
Moreover, the minimum is attained by a projection 
valued measure (PVM), where a PVM $\{E_{\alpha}\}$
is a family of the projection operators satisfying 
$\sum_{\alpha}E_{\alpha}={\bf 1}$.
\end{lemma}
\par
This lemma plays an essential role in the proof of our theorem.
The following theorem can be regarded as a 
generalized version of the Wigner-Araki-Yanase theorem on 
the distinguishability\cite{MIYAWAY}.
\begin{theorem}\label{maintheorem}
Let us consider a dynamics of the composite system, 
${\cal H}_A\otimes {\cal H}_B$, described by a unitary operator $U$.
Suppose that there exists a conserved quantity, $L:=L_A
+L_B+L_{int}$, where $L_A$ (resp.$L_B$) is 
an observable acting only on ${\cal H}_A$ (resp. ${\cal H}_B$),
and $L_{int}$ is an overlapping term.
The following inequality holds:
\begin{eqnarray}
|\langle \psi_0|L_A|\psi_1\rangle|
\leq \Vert L_B\Vert F(\rho_0^A,\rho_1^A)
+\Vert L_A \Vert F(\rho_0^B,\rho_1^B)
+2\Vert L_{int}\Vert,\label{tradeoff} 
\end{eqnarray}
where $\Vert \cdot \Vert$ is the operator norm 
defined as $\Vert v\Vert:=\sup_{|\phi\rangle \neq 0,|\phi\rangle\in {\cal H}}
\frac{\Vert v|\phi\rangle\Vert}{\Vert |\phi\rangle\Vert}$
for an operator $v$ on a Hilbert space ${\cal H}$.
\end{theorem}
{\bf Proof:}
By the purification of $\sigma$, we obtain 
a dilated Hilbert space and a vector state of the system $B$. 
We write the dilated Hilbert space as ${\cal H}_B$ for simplicity 
and the vector state as $|\Omega\rangle$. The dilated unitary 
operator $U \otimes {\bf 1}$ is also abbreviated as $U$.
Let us define initial vector states $|\Psi_i \rangle
:=|\psi_i\rangle \otimes |\Omega \rangle$ for $i=0,1$. 
Since $L$ is conserved with respect to the dynamics,
$ULU^* =L$ holds. Thus, 
as Wigner, Araki, and 
Yanase's original discussion\cite{Wigner,ArakiYanase,Yanase}, 
we have,
\begin{eqnarray}
\langle \Psi_0|L|\Psi_1 \rangle 
&=&
\langle \psi_0|L_A|\psi_1\rangle +\langle \Psi_0 |L_{int}|\Psi_1\rangle 
\nonumber \\
&=&
\langle \Psi_0|U^*(L_A+L_B+L_{int} )U|\Psi_1\rangle
\nonumber \\
&=&\langle \Psi_0|U^*L_AU|\Psi_1\rangle
+\langle \Psi_0|U^*L_BU|\Psi_1\rangle
+\langle \Psi_0|U^*L_{int}U|\Psi_1\rangle.
\label{fundamental}
\end{eqnarray}
Now we consider an arbitrary PVM $\{E_{\alpha}\}$ on the system $A$
and an arbitrary PVM $\{P_j\}$ on the system $B$.
Since $\sum_{\alpha}E_{\alpha}=\sum_{j}P_j={\bf 1}$ holds, 
the right hand side of (\ref{fundamental}) can be written as 
$\sum_{j} \langle \Psi_0|U^*P_j L_A U|\Psi_1\rangle
+\sum_{\alpha} \langle \Psi_0|U^*E_{\alpha} L_B U|\Psi_1\rangle
+\langle \Psi_0|U^* L_{int} U|\Psi_1 \rangle$.
By using commutativity $[P_j,L_A]=[E_{\alpha},L_B]=0$, 
we obtain,
\begin{eqnarray*}
\langle \psi_0|L_A|\psi_1\rangle
+\langle \Psi_0|L_{int}|\Psi_1\rangle
&=&
\sum_{j}\langle \Psi_0|U^*P_j L_A P_j U|\Psi_1\rangle
\\
&+&
\sum_{\alpha} \langle \Psi_0|U^*E_{\alpha} L_B E_{\alpha}
U|\Psi_1\rangle
+\langle \Psi_0|U^* L_{int} U|\Psi_1\rangle
.
\end{eqnarray*}
Taking absolute value of both sides, we obtain, 
\begin{eqnarray*}
|\langle \psi_0|L_A|\psi_1\rangle
+\langle \Psi_0|L_{int}|\Psi_1 \rangle
|&\leq &
\sum_{j}|\langle \Psi_0|U^*P_j L_A P_j U|\Psi_1\rangle|
+\sum_{\alpha}|\langle \Psi_0|U^*E_{\alpha} L_B E_{\alpha}
U|\Psi_1\rangle|
\\
&+&|\langle \Psi_0|U^* L_{int}U |\Psi_1\rangle|
\\
&\leq& \Vert L_A\Vert 
\sum_j \sqrt{\langle \Psi_0|U^*P_j U|\Psi_0\rangle
\langle \Psi_1|U^*P_j U|\Psi_1\rangle}
\\
&+&
\Vert L_B\Vert \sum_{\alpha}\sqrt{
\langle \Psi_0|U^*E_{\alpha} U|\Psi_0\rangle
\langle \Psi_1 |U^* E_{\alpha} U|\Psi_1\rangle}
+\Vert L_{int} \Vert.
\end{eqnarray*}
We here choose the particular PVMs, $\{E_{\alpha}\}$ and $\{P_j\}$, which 
attain the fidelity. Thanks to the lemma \ref{lemma1}
and the triangular inequality, we obtain,
\begin{eqnarray*}
|\langle \psi_0|L_A|\psi_1\rangle|
-|\langle \Psi_0|L_{int}|\Psi_1\rangle|
\leq
\Vert L_B\Vert F(\rho_0^{A},\rho_1^{A})
+\Vert L_A\Vert F(\rho_0^B,\rho_1^B)+\Vert L_{int}\Vert.
\end{eqnarray*}
Thus we obtain,
\begin{eqnarray*}
|\langle \psi_0|L_A|\psi_1\rangle|
\leq \Vert L_B\Vert F(\rho_0^{A},\rho_1^{A})
+\Vert L_A\Vert F(\rho_0^B,\rho_1^B)+2\Vert L_{int}\Vert.
\end{eqnarray*}
It ends the proof.
\hfill Q.E.D.
\par
The left hand side of the equation (\ref{tradeoff}) 
is related with noncommutativity between $L_A$ and 
the density operators $E_0:=|\psi_0 \rangle \langle \psi_0 |$ and 
$E_1:=|\psi_1\rangle \psi_1|$. 
In fact, if we put $N_j:=[E_j,L_A]\ (j=0,1)$, 
$|\langle \psi_0|L_A|\psi_1\rangle|^2
=\mbox{tr}(E_0 L_A E_1 L_A E_0)$ 
is expressed as:
\begin{eqnarray*}
\mbox{tr}(E_0 L_A E_1 L_A E_0)
&=&\mbox{tr}((L_A E_0 +N_0)E_1L_A E_0
)
\\
&=&
\mbox{tr}(N_0 E_1 L_A E_0)
\\
&=&
\mbox{tr}(N_0 (L_A E_1 +N_1)E_0)
\\
&=&
\mbox{tr}(N_0 N_1 E_0) =\langle \psi_0 |N_0 N_1|\psi_0\rangle.
\end{eqnarray*}
In contrast with the discussions on the Wigner-Araki-Yanase theorem
so far\cite{Wigner,ArakiYanase,Yanase,Ozawa1}, 
the conserved quantity is not restricted to be 
the additive one. Therefore we can treat the Hamiltonian itself as 
the conserved quantity. That is, we put $L_A=H_A$, $L_B=H_B$, and 
$L_{int}=H_{int}$.
\begin{theorem}\label{energymain}
Let us consider a composite system ${\cal H}_A\otimes {\cal H}_B$ 
that evolves by a Hamiltonian $H=H_A+H_B+H_{int}$ for 
an arbitrary time interval.
Suppose that an information bit is encoded to 
a pair of orthogonal states, $|\psi_0\rangle$ and $|\psi_1\rangle$, 
of the system $A$. For any choice of the initial state of 
the system $B$ that is independent of the bit encoded, 
the following inequalilty holds,
\begin{eqnarray}
|\langle \psi_0|H_A|\psi_1\rangle|
\leq \Vert H_B\Vert F(\rho_0^A,\rho_1^A)
+\Vert H_A\Vert F(\rho_0^B,\rho_1^B)
+2\Vert H_{int}\Vert,
\label{tradeoff2} 
\end{eqnarray}
where $\rho_j^A$ is the final states of system $A$ 
and $\rho_j^B$ is the final states of the system $B$
for the encoded bit $j$.
\end{theorem}
{\bf Proof:}
It is a direct consequence of the theorem \ref{maintheorem}.
\hfill Q.E.D.
\par 
Thus we obtain the strength of the interaction 
that is required for the perfect information distribution. 
\begin{corollary}
Under the condition of the theorem \ref{energymain},
if $2 \Vert H_{int}\Vert < |\langle \psi_0|H_A|\psi_1\rangle|$ holds,
the perfect information distribution cannot be attained.
\end{corollary}
{\bf Proof:}\\
The vanishing fidelities in (\ref{tradeoff2}) 
contradict with the nonvanishing left hand side.
\hfill Q.E.D.
\par
This corollary provides us a 
kind of no-go theorems. That is, even classical information 
cannot be copied if the strength of the interaction is 
smaller than a quantity determined by noncommutativity 
between the system Hamiltonian and the states (density operators) 
used for the information 
encoding.  
On the other hand, in case an approximate information distribution suffices, 
large systems can compensate the weakness of the interaction. 
\par
Let us consider the simplest example. The system $A$ 
is a spin $1/2$ system. 
The Hamiltonian of the system $A$ is the 
$z$-component of the spin, $H_A=S_z$. 
which 
is written with the 
eigenvectors, $|1\rangle$ and $|-1\rangle$, as 
$S_z=\frac{1}{2}(|1\rangle\langle 1|-|-1\rangle \langle -1|)$.
A pair of the orthogonal normalized vectors, $|\psi_1\rangle, 
|\psi_0\rangle$ can be written as 
$|\psi_1\rangle:=\alpha|1\rangle+\beta|-1\rangle$ and 
$|\psi_0\rangle:=\overline{\beta}|1\rangle-\overline{\alpha}|-1\rangle$
in general with neglect of irrelevant phase, 
where $|\alpha|^2+|\beta|^2=1$ is satisfied. 
The noncommutativity between these encoded states 
(density operators)
and $H_A$ can be calculated as,
$
\langle \psi_0|H_A|\psi_1\rangle =\alpha \beta.
$
Thus, to achieve the perfect information distribution, 
the strength of the interaction must satisfy, 
$\Vert H_{int}\Vert \geq |\alpha||\beta|/2$ 
which is nonvanishing in case $\alpha\neq 0$ 
and $\beta \neq 0$ hold. 
On the other hand, if $\alpha=1, \beta=0$ holds, the term 
$\langle \psi_0|H_A|\psi_1\rangle$ vanishes. 
In such a case, one can construct 
whatever weak interactions 
to achieve perfect 
information distribution.
Let us take the system $B$ as a spin $1/2$ system. 
If we put the initial state of the 
system $B$ as $|\Omega\rangle :=\frac{1}{\sqrt{2}}
(|1\rangle +|-1\rangle)$ and the Hamiltonian as $H_B={\bf 1}$ and 
$
H_{int}:=\epsilon(|1\rangle \langle 1|\otimes |1\rangle\langle 1|
+|-1\rangle \langle -1|\otimes |-1\rangle \langle -1|)$ for $\epsilon>0$,
$|1\rangle \otimes |\Omega\rangle$ evolves into 
$|1\rangle \otimes \frac{1}{\sqrt{2}}\left(
|1\rangle +e^{i \epsilon T}|-1\rangle 
\right)$ in time $T$, 
and 
$|-1\rangle \otimes |\Omega\rangle$ evolves into 
$|-1\rangle \otimes \frac{1}{\sqrt{2}}\left(
|1\rangle +e^{-i \epsilon T}|-1\rangle 
\right)$, where we neglected the phase.
Thus in time $T=\frac{\pi}{2\epsilon}$, 
we achieve the perfect information distribution. 
Since $\Vert H_{int}\Vert = \epsilon$ holds and $\epsilon >0$ is 
arbitrary, the strength of the interaction can be 
arbitrarily small.
\\
{\bf Acknowledgments:} We would like to thank the anonymous referee 
for helpful comments.

\end{document}